\documentstyle[amsmath,amssymb,graphicx]{article}

\newdimen\linethick  \linethick=0.4pt
\newdimen\hboxitspace    \hboxitspace=5pt
\newdimen\vboxitspace    \vboxitspace=5pt
\def\fr#1{%
\beq%\new
\vcenter{
\hrule height\linethick
          \hbox{\vrule width\linethick
                \kern\hboxitspace
                \vbox{\kern\vboxitspace
                      \hbox{$\begin{array}{c}\displaystyle#1
         \end{array}$}%
                      \kern\vboxitspace}%
                \kern\hboxitspace
                \vrule width\linethick}%
          \hrule height\linethick}%
\eeq}

\def\be{\begin{eqnarray}}
\def\ee{\end{eqnarray}}
\def\nn{\nonumber}

\def\beq{\begin{equation}}
\def\eeq{\end{equation}}

\def\p{\partial}

\def\PsiU{\Psi^{\rm U}}
\hoffset= - 1.0in         % switch off for draft style
\title{{\bf Link polynomial calculus and the AENV conjecture} \vspace{.2cm}}
\author{{\bf S.Arthamonov}\thanks{{\small
{\it ITEP, Moscow, Russia and Rutgers University, New Brunswick, USA}};
artamonov@itep.ru}, {\bf A.Mironov}\footnote{ {\small {\it Lebedev
Physics Institute} and {\it ITEP, Moscow, Russia}}; mironov@itep.ru;
mironov@lpi.ru}, {\bf A.Morozov}\thanks{{\small {\it ITEP, Moscow,
Russia}}; morozov@itep.ru}, \ {\bf And.Morozov}\thanks{{\small {\it
Moscow State University} and {\it ITEP, Moscow, Russia} and {\it Laboratory of Quantum Topology,
Chelyabinsk State University, Chelyabinsk, Russia}};
Andrey.Morozov@itep.ru}
\date{ }}

\begin{document}
 \maketitle

\vspace{-5.0cm}

\begin{center}
\hfill FIAN/TD-13/13\\
\hfill ITEP/TH-37/13\\
\end{center}

\vspace{3.5cm}

\centerline{ABSTRACT}

\bigskip

{\footnotesize
Using the recently proposed differential hierarchy (Z-expansion)
technique, we obtain a general expression for the HOMFLY polynomials
in two arbitrary symmetric representations
of link families, including Whitehead and Borromean links.
Among other things,
this allows us to check and confirm the recent conjecture of \cite{AENV}
that the large representation limit (the same as considered in the
knot volume conjecture) of this quantity matches the prediction
from mirror symmetry consideration. We also provide, using the evolution method, the HOMFLY polynomial
in two arbitrary symmetric representations
for an arbitrary member of the one-parametric family of
$2$-component $3$-strand links, which includes the Hopf and Whitehead links.
}

\vspace{3cm}

In \cite{AENV} the mirror symmetry was extended to include
augmentation varieties of links.
This allowed the authors to make a non-trivial prediction about the
Ooguri-Vafa partition function of the Whitehead link
in the volume conjecture limit \cite{vcl}.
As rightly noted in \cite{AENV}, the parallel recent
progress\footnote{
This progress is achieved on the route proposed
long ago by the seminal papers
\cite{ASch}-\cite{TR}.
}
in knot/link polynomial calculus in
\cite{progfirst}-\cite{AnoMcabling} would
allow one to test this conjecture by comparison with exact
formulas for the corresponding HOMFLY polynomials.
In this paper we make this comparison and indeed
confirm the AENV conjecture.

\section{Calculations of colored HOMFLY polynomials}

In this paper, results are reported only
for symmetric representations $S^r$, $S^s$ and $S^t$,
since only these are relevant for comparison with \cite{AENV}.

We used two ways to calculate the answers for the HOMFLY polynomials
and two ways to represent them.

Calculations are based on the ${\cal R}$-matrix approach \cite{TR,MMMkn12}
and use either the cabling method {\it a la} \cite{AnoMMMpaths,AnoMcabling}
or the eigenvalue hypothesis of \cite{IMMMev}.
The latter method is technically much simpler, but currently applicable
only to the case of $r=s=t$, hence, it was mainly used to check the results
of the former one.

The answers are represented either in the differential hierarchy ($Z$-expansion)
form of \cite{IMMMfe,Art}, which is convenient to control the
representation dependence, or in the evolution based form of \cite{DMMSS,evo},
convenient to control the dependence on the shape of the knot.
These two representations look very different, but in every particular case
one can easily convert between them.
Also both of them are useful for transition to superpolynomials.

In practice, the calculations were performed for a few low values of
$r,s,t$. Using the modern version of cabling approach, developed in
\cite{AnoMMMpaths,AnoMcabling} on the base of \cite{MMMkn12}, and
ordinary computers, one can handle up to 12-strand braids,
which means $r+s+t\leq 12$ for the three strand knots and links,
while the eigenvalue method of \cite{IMMMev} allows us to reach
the level of $r=s=t=4$ easier. After getting these explicit formulas, we
converted them into a differential expansion form {\it a la}
\cite{IMMMfe,Art}. As usual, {\it such} formulas in symmetric
representations have a pronounced q-hypergeometric form (in
accordance with a generic statement of \cite{Gar}) and are easily
continued to arbitrary values of $r$, $s$ and $t$.

%\begin{picture}(300,40)(-150,-15)
\begin{figure}
\begin{picture}(360,150)(-240,-50)
%\qbezier(0,0)(0,20)(20,20)
%\qbezier(0,0)(0,-20)(20,-20)
%%%-------Hopf-------%%%%
\qbezier(-160,0)(-160,10)(-150,17)
\qbezier(-160,0)(-160,-10)(-150,-17)
\qbezier(-150,17)(-140,25)(-132,19)
\qbezier(-150,-17)(-140,-25)(-130,-16)
\qbezier(-130,-16)(-110,0)(-128,16)
\qbezier(-101,0)(-101,10)(-111,17)
\qbezier(-101,0)(-101,-10)(-111,-17)
\qbezier(-111,-17)(-121,-25)(-129,-19)
\qbezier(-111,17)(-121,25)(-131,16)
\qbezier(-131,16)(-151,0)(-133,-16)
%%%%-------Whitehead-------%%%%
\put(-20,-20){\line(1,1){40}}
\put(2,-2){\line(1,-1){22}}
\put(-2,2){\line(-1,1){22}}
\qbezier(22,22)(44,0)(24,-20)
\qbezier(-22,-22)(-44,0)(-24,20)
\qbezier(22,22)(0,44)(-20,24)
\qbezier(-22,-22)(0,-44)(20,-24)
\qbezier(24,24)(45,45)(65,15)
\qbezier(24,-24)(45,-45)(65,-15)
\qbezier(65,15)(73,0)(65,-15)
\qbezier(-24,24)(-45,45)(-65,15)
\qbezier(-24,-24)(-45,-45)(-65,-15)
\qbezier(-65,15)(-73,0)(-65,-15)
%%%%-------Borromean-------%%%%
\qbezier(143,3)(155,19)(142,31)
\qbezier(140,0)(130,-10)(115,2)
\qbezier(138,35)(124,46)(112,30)
\qbezier(115,2)(100,12)(112,30)
\qbezier(145,16)(127,18)(120,2)
\qbezier(151,15)(163,13)(166,-6)
\qbezier(166,-6)(168,-30)(142,-30)
\qbezier(118,-2)(113,-30)(142,-30)
\qbezier(134,13)(141,-6)(163,0)
\qbezier(133,17)(130,26)(144,36)
\qbezier(144,36)(158,46)(170,32)
\qbezier(167,2)(187,9)(170,32)
\end{picture}
\caption{{\footnotesize These are three simplest links.
The leftmost figure shows the two-component Hopf link with the braidword $\sigma_1^2$.
The one in the middle is the two-component Whitehead link with the braidword
$\sigma_1\sigma_2^{-1}\sigma_1\sigma_2^{-1}\sigma_1$.
The rightmost link is the three-component Borromean rings link with the
braidword $\sigma_1\sigma_2^{-1}\sigma_1\sigma_2^{-1}\sigma_1\sigma_2^{-1}$}}
\end{figure}
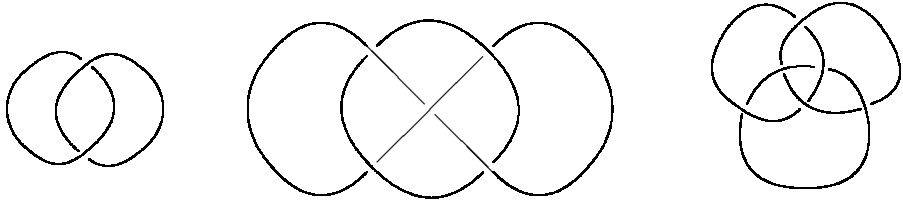

\newpage

In this way, we generalize the convenient expression for the Hopf link,
\be
\boxed{
H^{\rm H}_{r,s} = 1 + \sum_{k=1}^{{\rm min}(r,s)}(-1)^{k} A^{-k} q^{-k(r+s)+k(k+3)/2}
  \prod_{j=0}^{k-1}
 \frac{\{q^{r-j}\}\{q^{s-j}\}}{D_{j}}
}
\label{HopfHOMFLY}
\ee
to obtain formulas in the case of the Whitehead link,
\be
\boxed{
H_{r,s}^{\rm W}(A,q) = 1 + \sum_{k=1}^{{\rm min}(r,s)}\frac{1}{A^k q^{k(k-1)/2}}
\frac{D_{-1}}{D_{k-1}}\prod_{j=0}^{k-1}
\frac{D_{r+j}D_{s+j}}{D_{k+j}} \{q^{r-j}\}\{q^{s-j}\}
}
\label{WhHOMFLY}
\ee
and of the Borromean rings:
\be
\boxed{
H_{r,s,t}^{\rm B}(A,q) = 1 + D_{-1}\! \sum_{k=1}^{{\rm min}(r,s,t)}
(-)^k\{q\}^k \,[k]!\,
\frac{D_{k-2}!}{\big(D_{2k-1}!\big)^2}
\prod_{j=0}^{k-1} D_{r+j}D_{s+j}D_{t+j}
\{q^{r-j}\}\{q^{s-j}\}\{q^{t-j}\}
}
\label{BoHOMFLY}
\ee
Here $\{x\} = x-x^{-1}$ and $D_k = \{Aq^k\}$.
These expressions are manifestly symmetric under the permutation of
$r$ and $s$  or $r$, $s$ and $t$. These results did not appear in literature so far
\footnote{ We were
informed by S.Nawata that he is also aware of general formulas for
the Whitehead and Borromean links, and that they are consistent with
ours, see \cite{SN_Wh}. } and were tested at particular values of $r$, $s$ and $t$ indicated above
(the Whitehead HOMFLY polynomials were calculated up to $r+s\le 12$ and the Borromean ones up to $r+s+t\le 12$).

These answers are the {\it normalized}
HOMFLY, obtained by division over a product of two or three unknots
$S^*_rS^*_s$ or $S^*_rS^*_sS^*_t$, where the quantum dimension
\be
S^*_r = \prod_{i=0}^{r-1} \frac{D_i}{\{q^{r-i}\}}
\ee
Note that in the case of links (not knots), the normalized HOMFLY
is not a polynomial.
The first term is unity, as conjectured in \cite{evo},
this is a generic feature of
normalized knot polynomials in the topological framing.

In the paper, we also apply the evolution method of \cite{IMMMfe,evo} to the family of
$2$-component $3$-strand links (see Figure 2), which includes the Hopf link, the Whitehead link, L7a6, L9a36, L11a360
etc in accordance with the classification of \cite{katlas}. In this way, we obtain the HOMFLY polynomials for
this whole family in the case of the both symmetric representations, the result reads:
\fr{\label{5}
H^k_{r,s} = \sum_{p=0}^{{\rm min}(r,s)}  c^{(p)}_{r,s} q^{k\big((r-p)(s-p) - p\big)}\\
\\
c^{(p)}_{r,s} =
\frac{[r]![s]![r+s+1-2p]}{[p]![r+s+1-p]!}\cdot
\frac{D_{r+s-p-1}!  }{D_{r-1}!D_{s-1}!}\prod_{i=1}^{p} D_{i-2}\cdot
\left( 1+  \sum_{j=1}^{{\rm min}(r,s)} \frac{\sigma_{r,s}^{(p|j)}}{A^{2j}} \right)\\
\\
\sigma^{(p,j)}_{r,s}=
\displaystyle{{(-)^{p+j}\cdot q^{(p-1)(p-2j)}\cdot q^{\frac{j(j+1)}{2}-j(r+s)}\over [j]!}}\cdot
\{q\}^j\cdot \nn \\
\cdot
\sum_{a,b=0}^p\left(\frac{[p]!}{[a]![b]![p-a-b]!}\, (-)^{a+b}\cdot q^{(a+b)(j+1-p)}\cdot q^{ab}
\cdot\left( \prod_{i=0}^{j-1}\, [r-a-i][s-b-i]\right)\right)
}
This formula has no form of a $q$-hypergeometric polynomial. Therefore, the existing software
implementing Zeilberger's algorithm for the
hypergeometric sums \cite{Zeil,Koorw}
can not be immediately used to obtain the quantum $A$-polynomials.
For this purpose, the formula still has to be reshuffled like it is done in
the case of the Hopf (\ref{HopfHOMFLY}) and the Whitehead (\ref{WhHOMFLY}) links,
i.e. for $k=0$ and $k=2$ respectively.
In fact, revealing the differential hierarchy structure in this formula
for generic $k$ is a challenging problem of its own. We will return to this issue
elsewhere.

\section{Recursion relations for link polynomials}

Knot polynomials depend on various variables and satisfy various
interesting relations \cite{MirMoreqs}.
Of relevance for the AENV conjecture is dependence on the
"spins" $r,s,t$.
For links there are two kinds of such relations:
those relating evolution in $r$ and $s$ and involving the single
spin $r$ only. The former ones are very easy to observe.
The latter ones are usually complicated, but instead they can be
looked for in the same way as in the case of knots:
with the help of computer programs
implementing Zeilberger's algorithm for the
hypergeometric sums \cite{Zeil,Koorw}.
Recursion relations in $r$ are also sometime called
"quantum $A$-polynomials".

In the remaining part of this section we just list these
formulas for the Hopf and Whitehead links,
equations for the Borromean rings is too huge to be
included into the text.
Equations look a little simpler for non-normalized
polynomials ${\cal H}_{r,s,t} = S^*_rS^*_sS^*_t H_{r,s,t}$,
also it is {\it their} large representation asymptotic
that is to be compared with \cite{AENV}.

\subsection{Unknot}

The normalized HOMFLY polynomial for the unknot is just unity,
i.e. it does not depend on the representation,
\be
H^{\rm U}_r = H^{\rm U}_{r-1} = 1
\ee
However, for the non-normalized polynomial
${\cal H}^{\rm U}_r= S_r^*$ the same equation
looks already a little non-trivial:
\be
\{q^r\} {\cal H}^{\rm U}_r = D_{r-1} {\cal H}^{\rm U}_{r-1}
\label{recU}
\ee

For the purposes of the present paper it is useful
to consider the {\it restricted} Ooguri-Vafa (OV) partition
function which includes a sum over symmetric representations only:
we denote it by a bar. This restriction means that in the usual OV partition function
\be
{\cal Z}\{p_k\} = \sum_{R} {\cal H}_R S_R\{p_k\}
\ee
where $p_k=\sum_i z_i^k$ are auxiliary time variables (sources) and $S_R$ are the Schur functions,
one leaves only one Miwa variable $z_i=z$, i.e.
$p_k=z^k$. Then,
\be
\bar{\cal Z}(z) = \sum_{r} {\cal H}_r z^r
\ee
For the unknot all the four types of Ooguri-Vafa
functions are straightforwardly evaluated:
\be
{\cal Z}^{\rm U}\{p_k\} = \sum_{R} {\cal H}^{\rm U}_R S_R\{p_k\} =
\sum_R S_R^*S_R\{p_k\} = \exp\left(\sum_k \frac{p_k^*p_k}{k}\right), \nn \\
Z^{\rm U}\{p_k\} = \sum_{R} H^{\rm U}_R S_R\{p_k\} = \sum_R S_R\{p_k\} = e^{p_1}, \nn \\
\bar{\cal Z}^{\rm U}(z) = \sum_{r} {\cal H}^{\rm U}_r z^r =
\sum_r S_r^*z^r = \exp\left(\sum_k \frac{p_k^*z^k}{k}\right), \nn \\
\bar Z^{\rm U}(z) = \sum_{r} H^{\rm U}_r z^r = \sum_r z^r = \frac{1}{1-z}
\label{OVU}
\ee
Here we used the by-now-standard notation for the time variables
at the {\it topological locus}, where the Schur functions
turn into quantum dimensions, $S_R\{p_k^*\} = S_R^*$:
\be
p_k^* = \frac{\{A^k\}}{\{q^k\}} = \frac{A^k-A^{-k}}{q^k-q^{-k}} \
{\stackrel{A=q^N}{=\ }} \  \frac{[Nk]}{[k]}
\label{tolo}
\ee
The capital $R$ and small $r$ are used to denote {\it all} and only symmetric
representations respectively, the calligraphic ${\cal Z}$, ${\cal H}$
and ordinary letters $Z$, $H$ denote non-reduced and reduced
polynomials and partition functions.
The first two lines of (\ref{OVU}) are avatars of the Cauchy formula
for the Schur functions, while the last two are just direct corollaries
of definition of the Schur polynomials
\be
\sum_r S_r\{p_k\} z^k = \exp \left(\sum_k \frac{p_kz^k}{k}\right)
\label{Schurgen}
\ee
(note that sometime one use differently normalized time variables
$t_k = \frac{1}{k}p_k$, where this definition looks even simpler).
The partition function $\bar{\cal Z}^{|\rm U}(z)$ denoted through
$\PsiU(A|z)$ in \cite{AENV}
is just a further restriction of (\ref{Schurgen}) to the topological
locus (\ref{tolo}):
\be
\PsiU(A|z) = \exp \left(\sum_k \frac{p^*_kz^k}{k}\right)
= \exp\left(\sum_k \frac{(A^k-A^{-k})z^k}{k(q^k-q^{-k}}\right)
\label{OVu}
\ee
As a corollary of (\ref{recU}), it satisfies a difference equation in $z$.
Indeed, rewriting (\ref{recU}) as
\be
\sum_{r=0} z^{r+1} \Big(Aq^r - A^{-1}a^{-r}\Big)S_{[r]}^*
= \sum_{r=0} \Big(q^{r+1}-q^{-r-1}\Big) S_{[r+1]}^* z^{r+1}
\ee
one gets for the generating function
\be
z\Big(AT_z^+ - A^{-1}T_z^-\Big)\PsiU(A|z) =
\Big(T_z^+ - T_z^-\Big)\PsiU(A|z)
\ee
or
\be
\boxed{\ \
\Big( A\,(1-Aqz)\,T_z^2 + (A-qz)\Big) \,\PsiU(A|z) = 0\ \
\label{PsiUeq}
}
\ee
where the multiplicative-shift operators are defined by
$\ \hat T^{\pm}_z f(z) = f(q^{\pm 1}z)$.

An important additional observation is that the action
of the dilatation operator $T_A$
on the unknot function $\PsiU(A|z)$ is closely related to the action of
$T_z$:
\be
T^2_z \PsiU(A|z) = \PsiU(A|z)
\exp \left(\sum_k \frac{(A^k-A^{-k})(q^{2k}-1)z^k}{(q^k-q^{-k})k}\right)
=\nn \\
= \PsiU(A|z)\exp \left(\sum_k \frac{(A^k-A^{-k})\,q^kz^k}{k}\right)
= \frac{1-qz/A}{1-Aqz}\ \PsiU(A|z)
\ee
while
\be
T^2_A \PsiU(A|z) = \PsiU(A|z)
\exp \left(\sum_k \frac{\Big(A^k(q^{2k}-1)-A^{-k}(q^{-2k}-1)\Big)z^k}{(q^k-q^{-k})k}\right)
=\nn \\
= \PsiU(A|z)\exp \left(\sum_k \frac{\Big((Aq)^k+(Aq)^{-k}\Big)z^k}{k}\right)
= \frac{\PsiU(A|z)}{(1-Aqz)(1-z/Aq)}
\ee

\subsection{Hopf link}

There
are numerous different representations for the colored HOMFLY polynomial of the Hopf link
besides (\ref{HopfHOMFLY}).
To begin with, one could use the celebrated Rosso-Jones formula
\cite{RJ,china,BEM,DMMSS,MMMkn12}:
\be
{H}^{\rm H}_{R,S}\{p_k\} =
H_{R,S}^{\rm H}S_{R}\{p_k\}S_{S}\{p_k\} \sim q^{2\hat W\{p\}}
\Big(S_{R}\{p_k\}S_{S}\{p_k\}\Big)
\ee
where the cut-and-join operator
$\hat W = \hat W_{[2]} \sum_{a,b}\left((a+b)p_ap_b\frac{\p}{\p p_{a+b}}
+ abp_{a+b}\frac{\p^2}{\p p_a\p p_b}\right)$.

According to  \cite{AENV}, it is most convenient to begin from the
alternative representation for the Hopf link used in ref.\cite{AS}:
\be
{\cal H}^{\rm H}_{r,s} = \frac{S_{r}\{p_k^{(s)}\}}{S_{s}\{p_k^*\}} =
\frac{S_{s}\{p_k^{(r)}\}}{S_{r}\{p_k^*\}}
\label{ASHopf}
\ee
where the $r$-shifted topological locus is
\be
p_k^{(r)} = p_k^* - \frac{\{q^{kr}\}}{q^{k(r-1)}A^k}
= p_k^* - \frac{q^{kr}-q^{-kr}}{q^{k(r-1)}A^k}
\label{tolos}
\ee
Given these formulas, one can write \cite{AENV}
the restricted Ooguri-Vafa partition function for the Hopf link as
\be
\bar{\cal Z}^{\rm H}(x,y) =
\sum_{r,s} H^{\rm H}_{rs}\cdot S_{[r]}^*S_{[s]}^* \cdot x^r y^s
\ \stackrel{(\ref{ASHopf})}{=}\
\sum_{r,s}   S_{[r]}\{p_k^*\}  S_{[s]}\{p_k^{(r)}\} x^r y^s = \nn \\
\ \stackrel{(\ref{tolos})}{=}\ \sum_r S_{[r]}^* x^r
\exp\left( -\sum_k A^{-k}q^{-k(r-1)}(q^{kr} - q^{-kr})\frac{\p}{\p p_k}\right)
\sum_s S_{[s]}^* y^s = \nn \\
\ \stackrel{(\ref{Schurgen})}{=}\
\sum_r S_{[r]}^* x^r
\exp\left(-\sum_k {A^{-k}(q^{k} - q^{-k(2r-1)})}\frac{\p}{\p p_k}\right)
\left.\exp\left(\sum_k \frac{p_k y^k}{k}\right)\right|_{p_k=p_k^*} = \nn \\
=\sum_r S_{[r]}^* x^r
\exp\left(-\sum_k \frac{(q^{k} - q^{-k(2r-1)})y^k}{kA^k}\right)
\exp\left(\sum_k \frac{p_k^* y^k}{k}\right) = \nn \\
= \bar{\cal Z}^{\rm U}(y)\sum_r \frac{1-{qy}{A^{-1}}}{1-{y}{q^{-2r+1}A^{-1}}}\, S_{[r]}^* x^r
\ \stackrel{(\ref{Schurgen})}{=}\
\ \frac{1-qyA^{-1}}{1-qyA^{-1}\hat T_x^{-2}}\
\bar{\cal Z}^{\rm U}(x)\bar{\cal Z}^{\rm U}(y)
\ee
In other words,
\be
\boxed{
\bar{\cal Z}^{\rm H}(x,y) -
\frac{qy}{A}\bar{\cal Z}^{\rm H}(q^{-2}x,y)
= \left(1-\frac{qy}{A}\right)
\bar{\cal Z}^{\rm U}(x)\bar{\cal Z}^{\rm U}(y)
}
\label{ZH}
\ee
what implies for the non-normalized HOMFLY polynomials:
\be
{\cal H}^{\rm H}_{r,s}  -
\frac{q^{1-2r}}{A}{\cal H}^{\rm H}_{r,s-1}
=  S_r^*\Big(S^*_s - \frac{q}{A} S^*_{s-1}\Big)
\label{recHu}
\ee
For the normalized polynomials one gets, after using (\ref{recU}):
\be
\boxed{
H^{\rm H}_{r,s}-1 = \frac{q\{q^s\}}{AD_{s-1}}
\Big(q^{-2r}H^{\rm H}_{r,s-1}-1\Big)
}
\label{recH}
\ee
Additional equations are obtained by the substitution $x\leftrightarrow y$,
i.e. $r\leftrightarrow s$.

Complementary to these recursions in $s$ and $r$ there
is a simple relation, which involves shifts in the both directions and can be checked from the manifest
expression for the HOMFLY polynomial:
\be
\boxed{
{\cal H}^{\rm H}_{r,s}- q^2\,{\cal H}^{\rm H}_{r-1,s-1} =
\frac{q}{A}\cdot\frac{A^2q^{2r+2s-2}-1}{q^{2r}-q^{2s}}
\Big({\cal H}^{\rm H}_{r-1, s}-{\cal H}^{\rm H}_{r, s-1}\Big)
= \frac{D_{r+s-1}}{\{q^{r-s}\}}
\Big({\cal H}^{\rm H}_{r-1, s}-{\cal H}^{\rm H}_{r, s-1}\Big)
}
\label{simpleH}
\ee
Note that it looks more concise when written in terms of the
non-normalized HOMFLY.
It turns out that (\ref{simpleH}) remains {\it almost} the same in the
case of more complicated links, hence we call it
"simple" relation.

\subsection{Hopf link recursions from eq.(\ref{HopfHOMFLY})}

Our main task is, however, to deduce recurrence relations from
still another representation of the HOMFLY polynomial for the
Hopf knot, that is, from (\ref{HopfHOMFLY})
\be
H^{\rm H}_{r,s}\ \stackrel{(\ref{HopfHOMFLY})}{=}\
1 + \sum_{k=1}^{{\rm min}(r,s)}(-1)^{k} A^{-k} q^{-k(r+s)+k(k+3)/2}
  \prod_{j=0}^{k-1}
 \frac{\{q^{r-j}\}\{q^{s-j}\}}{D_{j}}
\label{HopfHOMFLY1}
\ee
What follows from (\ref{HopfHOMFLY1}) just immediately by
changing summation variable from $k$ to $k-1$, is
\be
H^{\rm H}_{r,s}\big(A\,\big|\,q\big) \ =\  1\  -\  \frac{\{q^r\}\{q^s\}}{Aq^{r+s-2} D_0}
\cdot H^{\rm H}_{r-1,s-1}\big(Aq\,\big|\,q\big)
\label{recuHopfH}
\ee
which can also be rewritten in two other ways:
\be
H^{\rm H}_{r,s}(A)-1 = q^s\{q^s\}\left(1 - \frac{D(r)}{q^rD_0}H^{\rm H}_{r,s-1}(Aq)\right)
\ee
and
\be
H^{\rm H}_{r,s}(A)+\frac{Aq^{r+s}D_rD_s}{D_0}\cdot H^{\rm H}_{r,s}(Aq) = A^2q^{2(r+s)}
\ee
For each given $r\geq s$ these are finite recursions to  $H_{r-s,0}^{\rm H}=1$
(we remind that for a link of unknots this quantity is symmetric under
the permutation of $r$ and $s$).
As a corollary, the Ooguri-Vafa generating function
\be
\bar {\cal Z}^{\rm H}(A|x,y) = \sum_{r,s} H^{\rm H}_{r,s}S_{[r]}^*S_{[s]}^*x^r y^s
\ee
satisfies
\be
\boxed{\
\bar{\cal Z}^{\rm H}(A|x,y) + \frac{xyD_0}{A}\,
\bar{\cal Z}^{\rm H}\Big(qA\,\Big|\,\frac{x}{q},\frac{y}{q}\Big)
= \bar{\cal Z}^{\rm U}(A|x)\,\bar{\cal Z}^{\rm U}(A|y)
\ }
\label{recuHopf}
\ee
Eq.(\ref{recuHopf}) is a recursion in a more tricky sense than (\ref{recuHopfH}):
the Ooguri-Vafa functions are power series, i.e. the series with only non-negative powers of $x$ and $y$.
Then (\ref{recuHopf}) allows one to express the coefficients in front of a given power
through those at lower powers, and thus reconstruct the entire series.

The derivation of (\ref{recuHopf})
makes use of the identity
\be
\{q^r\}S_{[r]}^*(A|q) = D_0 S_{[r-1]}^*(Aq|q)
\ee
which is an important complement of (\ref{recU}).
Note that (\ref{recuHopfH}) is a similar  complement of (\ref{recH}).

However, it is (\ref{recuHopfH}) which is a {\it straightforward}
implication of (\ref{HopfHOMFLY1}): derivation of (\ref{recH})
from {\it this} starting point is somewhat more transcendental.
It uses the fact that the hypergeometric polynomials often satisfy
difference relations w.r.t. its {\it parameters}, not only {\it arguments},
and (\ref{HopfHOMFLY1}) is exactly of this type.
Indeed, it can be rewritten in terms of the $q$-factorials
$(n)! = \prod_{i=1}^n(1-q^{2i})$:
\be
H^{\rm H}_{r,s}(A=q^N) = \sum_{k=0} h_{r,s}(k) =
\sum_{k=0} q^{2k(k-r-s)}\frac{(r)!(s)!(N-1)!}{(r-k)!(s-k)!(N+k-1)!},\nn\\
{\cal H}^{\rm H}_{r,s}(A=q^N) = \sum_{k=0} {\mathfrak{h}}_{r,s}(k) =
\sum_{k=0} q^{2k(k-r-s)-(N-1)(s+r)}\frac{(N+r-1)!(N+s-1)!}{(r-k)!(s-k)!(N+k-1)!(N-1)!}
\ee
and acquires a form of the $q$-hypergeometric polynomials
(for the $q$-hypergeometric functions of type $_3F_1$,
note that the $(-)^k$ in this case is absorbed into
the $q$-factorials).
In this case, the recursion in the parameter $s$ can, of course,
be found "by hands", but this is almost impossible for
more general series of this type like (\ref{WhHOMFLY})
and (\ref{BoHOMFLY}).
Therefore, it makes sense to apply the standard software \cite{Zeils},
one should only divide the quadratic form in the exponent of $q$
by two, because the program uses $q$ instead of $q^2$.
The program gives the equations in the case of reduced and non-reduced polynomials in the form
\be\label{equnr}
q^{-2r}{q\over A}{\mathfrak{h}}_{r,s}(k)-{\mathfrak{h}}_{r,s+1}(k)={\cal G}_{r,s}(k+1)-{\cal G}_{r,s}(k)
\ee
and
\be\label{eqr}
h_{r,s}(k)-{A\over q}q^{2r}{\{Aq^s\}\over\{q^{s+1}\}}h_{r,s+1}(k)=G_{r,s}(k+1)-G_{r,s}(k)
\ee
where
\be
{\cal G}_{r,s}(k)\equiv q^{s-2k+1}{\{Aq^{k-1}\}\over\{q^{s-k+1}\}}{\mathfrak{h}}_{r,s}(k);\\
G_{r,s}(k)\equiv {\{Aq^{k-1}\}\over\{q^{s-k+1}\}}Aq^{2r+s-2k}h_{r,s}(k)
\ee
When we sum (\ref{equnr}) and (\ref{eqr}) over $k$ from 0 to infinity,
then only the lower summation limit at $k=0$ contributes at the r.h.s.,
because both $h(k)$ and $\mathfrak{h}(k)$ vanish at large $k>{\rm min}(r,s)$.
Since
\be
-{\cal G}_{r,s}(0)=S_r^*\Big(S^*_s - \frac{q}{A} S^*_{s-1}\Big)
\ee
and
\be
-G_{r,s}(0)=q^{2r}-Aq^{2r-1}{\{Aq^s\}\over\{q^{s+1}\}}
\ee
this immediately leads to (\ref{recHu}) and (\ref{recH}).

It deserves noting that instead of the first order difference
equations with a non-vanishing free term, one can  write down
a {\it homogeneous} equation of the second order in the shift operator:
\be\label{Hh}
\Big(q^{2s}-{q^4\over A^2}\Big){\cal H}^{\rm H}_{r,s-2}-{q\over A}
\Big(q^{2s}-1\Big){\cal H}^{\rm H}_{r,s-1} -{A\over q}\Big(q^{2s}
-{q^4\over A^2}\Big)q^{2r}{\cal H}^{\rm H}_{r,s-1}
+\Big(q^{2s}-1\Big)q^{2r}{\cal H}^{\rm H}_{r,s}=0
\ee
This follows directly from (\ref{recHu}) in the form
\be
\{q^s\}\Big({\cal H}_{r,s}^{\rm H} - \frac{q^{1-2r}}{A}{\cal H}_{r,s-1}^{\rm H}\Big)
=qD_{s-2}\Big({\cal H}_{r,s-1}^{\rm H} - \frac{q^{1-2r}}{A}{\cal H}_{r,s-2}^{\rm H}\Big)
\ee
Equation (\ref{Hh}) can be obtained by applying the operator annihilating the unknot $S^*_r$, (\ref{recU})
\be\label{Or}
\hat O_r\equiv \{q^r\}-D_{r-1}\hat P_r,\ \ \ \ \ \ \ \ \hat P_r {\cal H}_r\equiv {\cal H}_{r-1}
\ee
to (\ref{recHu}): since the r.h.s. of (\ref{recHu}) depends on $r$ only though $S^*_r$, one immediately
obtains a homogeneous equation (of the second order).

\subsection{Whitehead link}

Now we switch to the Whitehead link, with the HOMFLY polynomial
in symmetric representations given by (\ref{WhHOMFLY}):
\be
H_{r,s}^{\rm W}(A,q) = 1 + \sum_{k=1}^{{\rm min}(r,s)}\frac{1}{A^k q^{k(k-1)/2}}
\frac{D_{-1}}{D_{k-1}}\prod_{j=0}^{k-1}
\frac{D_{r+j}D_{s+j}}{D_{k+j}} \{q^{r-j}\}\{q^{s-j}\} =
\nn \\
= 1 + D_{-1}\!\sum_{k=1}^{{\rm min}(r,s)}\frac{1}{A^k q^{k(k-1)/2}}\frac{D_{k-2}!}{D_{2k-1}!}
 \prod_{j=0}^{k-1}
D_{r+j}D_{s+j}\, \{q^{r-j}\}\{q^{s-j}\}
\label{WhHOMFLY1}
\ee
We  are again interested in recurrence relations in $r$ and $s$.

First of all, one can check that it satisfies the "simple" relation,
which is practically the same as (\ref{simpleH}) for the Hopf link:
\be
\boxed{
{\cal H}^W_{r,s} - {\cal H}^W_{r-1,s-1} =
\frac{q}{A}\cdot\frac{A^2q^{2r+2s-2}-1}{q^{2r}-q^{2s}}
\Big({\cal H}^W_{r-1,s}-{\cal H}^W_{r,s-1}\Big)
= \frac{D_{r+s-1}}{\{q^{r-s}\}}
\Big({\cal H}^{\rm W}_{r-1, s}-{\cal W}^{\rm H}_{r, s-1}\Big)
}
\label{simpleW}
\ee
The only difference is that in (\ref{simpleH})
there is a factor $q^2$ in front of the second item at the l.h.s.
We remind that the HOMFLY polynomial in this formula is non-normalized,
${\cal H}^{\rm W}_{r,s} = H^{\rm W}_{r,s}S_r^*S_s^*$.

Second, the "natural" recursion  for
(\ref{WhHOMFLY}) is somewhat less trivial than (\ref{recuHopfH}):
$H^{\rm W}_{r,s}(A|q) = H^{(0)}_{r,s}(A|q)$ is just the first term in the additional hierarchy
\be
H^{(m)}_{r,s}(A|q) = 1 +
\frac{D_{m-1}D_{r+m}D_{s+m}}{q^mA D_{2m}D_{2m+1}}
\{q^{r-m}\}\{q^{s-m}\}\cdot H^{(m+1)}_{r,s}(A|q)
\label{recuWH}
\ee
With this definition, $H^{(m)}_{r,s}\neq 1$ only for $r,s>m$
and this relation is indeed a recursion with a finite number of steps
needed to find any particular term in these polynomials.
In accordance with (\ref{recuWH}), the Ooguri-Vafa partition function
\be
\bar{\cal Z}^{\rm W}(A|x,y) = \sum_{r,s} H^{\rm W}_{r,s}S_{[r]}^*S_{[s]}^*x^r y^s =
\bar{\cal Z}^{(0)}(A|x,y)
\ee
is just the first member of the hierarchy, and a direct counterpart of
(\ref{recuHopf}) is obtained when one multiplies (\ref{recuWH}) by
$S_{[r-m]}^*(q^{2m}A)\, S_{[s-m]}^*(q^{2m}A)$
and makes use of
$D_{r+m}\{q^{r-m}\} S_{[r-m]}^*\left(q^{2m}A\right)
= D_{2m}D_{2m+1} S_{[r-m-1]}^*\left(q^{2(m+1)}A\right)$:
\be
\bar{\cal Z}^{(m)}(A|x,y) \equiv
\sum_{r,s} H^{(m)}_{r,s}(A|q)S_{[r-m]}^*(q^{2m}A)S_{[s-m]}^*(q^{2m}A)
x^{r-m} y^{s-m} = \nn \\
\ \stackrel{(\ref{recuWH})}{=}\
\sum_{r,s}  S_{[r-m]}^*(q^{2m}A)S_{[s-m]}^*(q^{2m}A) x^{r-m} y^{s-m} + \nn \\
+ \ \frac{D_{m-1}D_{2m}D_{2m+1}}{q^mA}
\sum_{r,s} H^{(m+1)}_{r,s}(A|q)S_{[r-m-1]}^*\left(q^{2(m+1)}A\right)
S_{[s-m-1]}^*\left(q^{2(m+1)}A\right)x^{r-m}y^{s-m} \ \ \ \ \ \ \ \ \ \
\ee
i.e.
\be
\boxed{
\bar{\cal Z}^{(m)}(A|x,y) -
\frac{xy D_{m-1}D_{2m}D_{2m+1}}{q^mA}\,\bar{\cal Z}^{(m+1)}(A|x,y)
= \bar{\cal Z}^{\rm U}(Aq^{2m}|x)\,\bar{\cal Z}^{\rm U}(Aq^{2m}|y)
}
\label{recuW}
\ee
Eq.(\ref{simpleW}) can also be rewritten in terms of the OV partition function,
this time without any decomposition:
\be
\boxed{
(1-xyq^2)\Big(\bar{\cal Z}^{\rm W}(q^2x,y)-\bar{\cal Z}^{\rm W}(x,q^2y)\Big) =
\frac{q}{A}(x-y)\Big(A^2 \bar{\cal Z}^{\rm W}(q^2x,q^2y) - \bar{\cal Z}^{\rm W}(x,y)\Big)
}
\ee

Third, the most non-trivial recursion in $s$ only (with $r$ and $A$ fixed)
can be again obtained with the help of the program \cite{Zeils},
this time it should be applied to the $q$-hypergeometric polynomial
(\ref{WhHOMFLY1}):
$$
H^{\rm W}_{r,s}(A=q^N)=\sum_{k=0} (-)^{k+1} q^{k^2+k-2k(r+s+N)+N-1}\ D_{-1}
\frac{(r)!(s)!(r+N+k-1)!(s+N+k-1)!(N+k-2)!}{(r-k)!(s-k)!(r+N-1)!(s+N-1)!(N+2k-1)!}
$$
\vspace{-0.3cm}
\be
{\cal H}^{\rm W}_{r,s}(A=q^N) =
\sum_{k=0} (-)^{k+1} q^{k^2+k-2k(r+s+N)+(1-s-r)(N-1)}\ D_{-1} \times\nn\\ \times
\frac{((r+N+k-1)!(s+N+k-1)!(N+k-2)!}{(r-k)!(s-k)!(N+2k-1)!(N-1)!(N-1)!}\equiv \sum_k{\mathfrak{h}}^{\rm W}_{r,s}(k)
\ee
Similar to the Hopf case, one obtains for the non-normalized quantities the equation
\be
{D_{2s+5}\over A^2q^{2s}}{\mathfrak{h}}^{\rm W}_{r,s}(k)+\nn\\
+\left(\Big(q^{r-s+2}+q^{-r-s}-q^{s-r+4}\Big){D_{r+s+1}D_{2s+5}\over A^2}+
{D_{s+5}D_{2s+5}\over A^2q^s}-{q^{r-2s-3}\over A^2}D_{r+2}D_{2s+5}
-q^{2s+8}\{q^2\}\right){\mathfrak{h}}^{\rm W}_{r,s+1}(k)+\nn\\
\hspace{-1cm}+D_{2s+4}\left({q^3\over A^2}D_{s+r+2}^2+{q^{1-r}\over A^2}D_{s+r+2}(D_{s+2}+q^3D_{s+1})
-{q^{2-r}\over A^2}D_{2s+5}D_r-{q^{1-s}\over A^2}D_{2r-1}D_{s+1}-{q^{3-2r}\over A^2}(A^2q^{4r}+1)
\right){\mathfrak{h}}^{\rm W}_{r,s+2}(k)-\nn\\
-\left({q^{r+2}\over A}D_{2s+3}D_{s+2}D_{s+r+3}+{q^{s-r+6}\over A^2}D_{2s+3}D_{s+r+3}+
{q^{2s-2r+7}\over A}D_{2s+3}-{q^3\over A}D_{2s+5}\right){\mathfrak{h}}^{\rm W}_{r,s+3}(k)+\nn\\
+q^{8+2s}D_{2s+3}{\mathfrak{h}}^{\rm W}_{r,s+4}(k)={\cal G}_{r,s}(k+1)-{\cal G}_{r,s}(k)
\ee
where
\be
{\cal G}_{r,s}(k)=
q^{2k}D_{2k-1}D_{2k-2}D_{2s+5}D_{2s+4}D_{2s+3}\{q^{2-k}\}\{q^{3-k}\}
\{q^{4-k}\}\{q^{5-k}\}\prod_{i=0}^{s-2}{\{q^{k-2-i}\}\over\{q^{k-6-i}\}}{\mathfrak{h}}^{\rm W}_{r,s}(k)
\ee
so that
\be
-{\cal G}_{r,s}(0)=\frac{S_r^*D_{2s+3}D_{2s+4}D_{2s+5}}{ \{q^{s+3}\}\{q^{s+4}\}}
\prod_{i=1}^{s+2}{D_{i-3}\over  \{q^i\}}
\ee
and this leads to the equation
\be\label{WHcur}
\!\!\!\!\!
{D_{2s+5}\over A^2q^{2s}}{\cal H}^{\rm W}_{r,s}+\nn\\
+\left(\left(\Big(q^{r-s+2}+q^{-r-s}-q^{s-r+4}\Big)D_{r+s+1}+
q^{-s}D_{s+5}-q^{r-2s-3}D_{r+2}\right){D_{2s+5}\over A^2}\
\underline{-\ q^{2s+8}\{q^2\}}\right){\cal H}^{\rm W}_{r,s+1}+\nn\\
\hspace{-1cm}+\Big({q^3}D_{s+r+2}^2+{q^{1-r}}D_{s+r+2}(D_{s+2}+q^3D_{s+1})
-{q^{2-r}}D_{2s+5}D_r-{q^{1-s}}D_{2r-1}D_{s+1}-{q^{3-2r}}(A^2q^{4r}+1)
\Big) {D_{2s+4}\over A^2} {\cal H}^{\rm W}_{r,s+2}-\nn\\
-\left({q^{r+2}\over A}D_{2s+3}D_{s+2}D_{s+r+3}+{q^{s-r+6}\over A^2}D_{2s+3}D_{s+r+3}+
{q^{2s-2r+7}\over A}D_{2s+3}-{q^3\over A}D_{2s+5}\right){\cal H}^{\rm W}_{r,s+3}+\nn\\
+q^{8+2s}D_{2s+3}{\cal H}^{\rm W}_{r,s+4}= \nn
\ee
\vspace{-0.3cm}
\be
= \frac{S_r^*D_{2s+3}D_{2s+4}D_{2s+5}}{ \{q^{s+3}\}\{q^{s+4}\}}
\prod_{i=1}^{s+2}{D_{i-3}\over  \{q^i\}}
\ee
As in the Hopf case, the r.h.s. of this equation depends on $r$ only though the unknot $S^*_r$, i.e.
one can again apply the operator $\hat O_r$ (\ref{Or}) in order to get a (fifth order) homogeneous equation.

\subsection{Borromean rings}

For the Borromean 3-component link, the normalized HOMFLY polynomial is
given by (\ref{BoHOMFLY}):
\be
H_{r,s,t}^{\rm B}(A,q) = 1 + D_{-1}\! \sum_{k=1}^{{\rm min}(r,s,t)}
(-)^k\{q\}^k \,[k]!\,
\frac{D_{k-2}!}{\big(D_{2k-1}!\big)^2}
\prod_{j=0}^{k-1} D_{r+j}D_{s+j}D_{t+j}
\{q^{r-j}\}\{q^{s-j}\}\{q^{t-j}\}
\label{BoHOMFLY1}
\ee
Note that the corrections to $1$ are of order $\{q\}^4$.
Also for $r$ or $s$ or $t=[0]$ the answer is just unity:
because when one component of the Borromean link is removed,
the other two are two independent unknots.

The "simple" relation for non-normalized ${\cal H}^B$
is literally the same as (\ref{simpleW}):
\be
\boxed{
{\cal H}^B_{r,s,t}- {\cal H}^B_{r-1,s-1,t} =
\frac{q}{A}\cdot\frac{A^2q^{2r+2s-2}-1}{q^{2r}-q^{2s}}
\Big({\cal H}^B_{r-1, s, t}-{\cal H}^B_{r, s-1, t}\Big)
= \frac{D_{r+s-1}}{\{q^{r-s}\}}
\Big({\cal H}^{\rm B}_{r-1, s}-{\cal H}^{\rm B}_{r, s-1}\Big)
}
\label{simpleB}
\ee
Of course, this time there are two more relations for the
pairs $r,t$ and $s,t$.

Rewriting (\ref{BoHOMFLY1}) as a hypergeometric polynomial:
\be
H^{\rm B}_{r,s,t}(A=q^N) -1 =
\sum_k (-)^k q^{3k^2-k-2k(r+s+t)+\frac{N(N+1)}{2}-1}\times \nn \\ \times
\frac{(r)!(s)!(r+N+k-1)!(s+N+k-1)!(t+N+k-1)!(N+k-2)!k!}{(r-k)!(s-k)!(t-k)!(r+N-1)!(s+N-1)!
(t+N-1)!\left((N+2k-1)!\right)^2}
\ee
and in non-normalized case as
\be
{\cal H}^{\rm B}_{r,s,t}(A=q^N) =
\sum_k (-)^k q^{3k^2-k-2k(r+s+t)+(1-N)(s+r+t)+\frac{N(N+1)}{2}-1}\times \nn \\ \times
\frac{(r+N+k-1)!(s+N+k-1)!(t+N+k-1)!(N+k-2)!k!}{(r-k)!(s-k)!(t-k)!\left((N+2k-1)!\right)^2}
\ee
we can apply the program of \cite{Zeils}
to get the recursion relation in $s$ (or $r$ or $t$),
which is now an order six difference equation.
It is, however, too huge to be presented here --
but can be easily generated by MAPLE or Mathematica.
For illustrative purposes we present just the two simplest items
in the equation for non-normalized HOMFLY polynomial:
\be
D_{2s+3}D_{2s+4}D_{2s+5}D_{2s+10}D_{2s+11}\,{\cal H}^{\rm B}_{r,s,t}
+ \ldots +
D_{2s+1}D_{2s+2}D_{2s+7}D_{2s+8}D_{2s+9}\, {\cal H}^{\rm B}_{r,s+6,t}
=  \ldots
\ee

\subsection{More three-component links}

In paper \cite{AENV} there are three more three-component links considered. These are: the (3,3)-torus link,
a connected sum of two Hopf links (the composite Hopf link) and link $L8n5$ in accordance with the
Thistlethwaite Link Table \cite{katlas}. We omit here the torus link, since all the formulas in this case are
immediate. Another considered pattern, the composite Hopf link is also simple, since its normalized HOMFLY
polynomial satisfies (like any composite link)
\be
H_{r_1,r_2,r_3} = H_{r_1,r_2}^{\rm H} H_{r_2,r_3}^{\rm H}
\ee
The braidword of this link is $\sigma_1\sigma_1\sigma_2\sigma_2$.

The least trivial example is link $L8n5$.
The braidword of this link is $\sigma_1\sigma_2^{-1}\sigma_1\sigma_2^{-1}\sigma_1(\sigma_2)^3$.
It can be constructed from the Borromean rings by adding four inverse crossings, i.e. it
belongs to the same evolution series with k=-4, see section 5.

\section{Spectral curves}

The linear recurrence relations in the previous section are actually
written in terms of two operators
$\hat Q_i$ and $\hat P_i$, which act on the representation index
of HOMFLY polynomials as follows:
\be
\hat Q_r {\cal H}_r = q^{r} {\cal H}_r, \nn \\
\hat P_r {\cal H}_{r} = {\cal H}_{r-1}
\ee
These operators satisfy the commutation relation
\be
\hat P_i\hat Q_j = q^{\delta_{ij}} \hat Q_j \hat P_i
\ee
and they commute in the limit $q = e^{\hbar}\longrightarrow 1$.
Therefore, in this limit in an appropriate basis,
which is in fact provided by the (restricted) Ooguri-Vafa functions,
one can substitute the difference equations by
a vanishing condition for a system of polynomials.
All together they define an algebraic variety,
which is called the {\it spectral variety},
associated with the given link or knot.
Moreover, this spectral variety is known to coincide
with the classical ${\cal A}$-polynomial.

The AENV conjecture \cite{AENV} is about these classical $A$-polynomials,
which are independently calculated by topological methods,
and our purpose in this section is to demonstrate that they
are indeed obtained as the spectral curves for our
knot polynomials.
This check proves the AENV conjecture for the Whitehead and
the Borromean links.

\subsection{Unknot}

At small $\hbar$ ($q=e^\hbar$) the Ooguri-Vafa function (\ref{OVu}) behaves as follows
\be
\bar{\cal Z}^{\rm U} =
\exp\left(\sum_k \frac{p_k^*z^k}{k}\right)
%=\nn \\
= \exp\left(\sum_k \frac{(A^k-A^{-k})z^k}{k(q^k-q^{-k})}\right)
= \exp \left(\frac{1}{\hbar}W^{\rm U}(A|z) + O(1)\right)
\ee
and the genus zero free energy of the unknot is
\be
W^{\rm U}(A|z) = \sum_k \left(\frac{A^kz^k}{2k^2} - \frac{A^{-k}z^k}{2k^2}\right)
\label{Wun}
\ee
The spectral curve is the {\it algebraic}
relation between $\mu = \exp\left(z\frac{\p W(A|z) }{\p z}\right)$ and $z$.
From (\ref{Wun})
\be
2\log \mu =  \sum_k \left(\frac{A^kz^k}{k}-\frac{A^{-k}z^k}{k}\right)
= \log(1-z/A)-\log(1-zA)
\ee
and one obtains the spectral curve  for the unknot:
\be
\boxed{\ \Sigma^{\rm U}:\ \ \ \ \ \ \ \ \ \mu^2\,(1-zA) = 1-z/A\ }
\label{Sigmaunknot}
\ee
with the Seiberg-Witten differential $\log \mu\frac{dz}{z}$.
Changing the variables,
\be
x=\frac{1}{\mu^{2}},\ \ \ \ \  y=\frac{A}{z}
\label{chava}
\ee
one can rewrite it as \cite{AENV}:
\be\label{uncur}
(A/z) + \mu^{-2} - (A/z)\mu^{-2} = A^2 \ \
\longrightarrow \ \
\boxed{\ x + y -xy = A^2\ }
\ee

Equivalently one can obtain the same spectral curve (\ref{Sigmaunknot})
as the $q=1$ limit  of (\ref{PsiUeq}),
provided the action of the dilatation operator $\hat T$
is substituted by the quasi-momentum $\mu$:
\be
%\boxed{\ \
\Big( A\,(1-Aqz)\,T_z^2 + (A-qz)\Big) \,\PsiU(A|z) = 0\ \
\stackrel{q\rightarrow 1}{\longrightarrow} \
(1-zA)\mu^2 = 1-z/A
\label{PsiUeqsc}
%}
\ee

Useful in applications is also the expansion
\be
S_{[r]}^* = \prod_{j=0}^{r-1} \frac{D_j}{\{q^{r-j}\}} \sim
\frac {\{A\}^r}{(2\hbar)^r r!}\left( 1
+ \hbar\cdot \sum_{j=0}^{r-1} \frac{j(A+A^{-1})}{A-A^{-1}}
+ \ldots \right)
%= \nn \\
= \frac {\{A\}^r}{(2\hbar)^r r!}
+ \frac{\{A^2\}}{8\hbar} \cdot \frac {\{A\}^{r-2}}{(2\hbar)^{r-2} (r-2)!} + \ldots
\ee

\bigskip

Still for our purposes it is desirable to derive $\Sigma^{\rm U}$ directly from
(\ref{recU}).
This is, of course, straightforward:
\be
\{q^r\} {\cal H}^{\rm U}_r = D_{r-1} {\cal H}^{\rm U}_{r-1}
\ \ \ \Leftrightarrow \ \ \
\left\{\left(\hat Q - \frac{1}{\hat Q}\right) - \left(A\hat Q - \frac{1}{A\hat Q}\right)
\right\} {\cal H}_r^{\rm U} = 0
\ee
or
\be\label{canun}
\left\{\hat Q^2(1-A\hat P) - (1-A^{-1}\hat P)\right\}{\cal H}_r^{\rm U} = 0
\ee
Since for the generating OV function $\Psi(z) = \sum_r {\cal H}_rz^r$
\be
\hat P \Psi(z|A) =z\Psi(z|A), \nn \\
\hat Q \Psi(z|A) = \hat T_+ \Psi(z|A)
\ee
this equation is just the same as (\ref{PsiUeqsc}).

\subsection{Hopf link}

In the small $\hbar$ limit of (\ref{recuHopf}),
\be
\bar{\cal Z}^{\rm H}(A|x,y) + \frac{xyD_0}{A}\,
\bar{\cal Z}^{\rm H}\Big(qA\,\Big|\,\frac{x}{q},\frac{y}{q}\Big)
= \bar{\cal Z}^{\rm U}(A|x)\,\bar{\cal Z}^{\rm U}(A|y)
\ee
there are two possibilities:
one is that $Z^{\rm Hopf}$ is not sensitive to the small (by $\hbar$) shift
of its parameters, i.e. is regular in the limit $q\to 1$; then
\be\label{75}
{\rm phase} \ 1\otimes 1 \ {\rm of} \ \cite{AENV}:
\ \ \ \ \ \
\bar{\cal Z}_{(1\otimes 1)}^{\rm H}(A|x,y) \sim
\frac{\bar{\cal Z}^{\rm U}(A|x)\bar{\cal Z}^{\rm U}(A|y)}
{1+\frac{xyD_0}{A}},\nn \\ \nn \\
{\rm i.e.} \ \ \ \ \ \
W^{\rm H}_{(1\otimes 1)}(A|x,y) = W^{\rm U}(A|x)+ W^{\rm U}(A|y)
\ee

However, there is still a possibility to add to
$\bar{\cal Z}_{(1\otimes 1)}^{\rm H}(A|x,y)$
a solution of the {\it homogeneous} equation at the l.h.s. of (\ref{recuHopf}):
\be
{\rm phase} \ 2 \ {\rm of} \ \cite{AENV}:
\ \ \ \ \ \
\bar{\cal Z}^{\rm H}_{(2)}(A|x,y) =
- \frac{xyD_0}{A}\bar{\cal Z}_{(2)}^{\rm H}\Big(qA\Big|\frac{x}{q},\frac{y}{q}\Big)
\label{pha2}
\ee
This equation has a factorized solution
\be
\bar{\cal Z}_{(2)}^{\rm H}(A|x,y) = f(A)Z(x,y)
\ee
with
\be
Z(x,y) = xy\cdot Z\Big(\frac{x}{q},\frac{y}{q}\Big) &
\Longrightarrow &  Z(x,y) \sim \exp\left(\frac{\log x\log y}{\log q}\right)
= \exp \left(\frac{\xi\eta}{\hbar}\right),  \\ \nn \\
f(A) = (A^{-2}-1)f(Aq) & \Longrightarrow &  f(A) = N\prod_{i=0}^\infty
\Big((q^iA)^{-2}-1\Big)^{-1} \nn
\ee
where $N$ is an arbitrary constant. This "homogeneous part" of solution
is singular at the point $q=1$ and, hence, predominates
over (\ref{75}).

This solution is just the one suggested in \cite{AENV}
to describe the non-trivial phase $"2"$ of the Hopf link OV partition
function:
the double Fourier transform with the weight $\ \exp\left(\frac{ip\xi+ip'\eta}{\hbar}\right)\ $
is
\be
\hat {\cal Z}^{\rm H}_{(2)}(A|p,p') = f(A) \exp\left(\frac{pp'}{\hbar}\right)
\ee
This function before and after the Fourier transform satisfies the
peculiarly simple differential equations like
\be
\left(\frac{\p}{\p p} - p'\right)\hat {\cal Z}^{\rm H}_{(2)}(A|p,p') =
\left(\frac{\p}{\p p'} - p\right)\hat {\cal Z}^{\rm H}_{(2)}(A|p,p') = 0
\ee

Our next task is to derive them directly from the recurrence relation
(\ref{recHu}). This is straightforward.
In the phase $2$ of \cite{AENV},
when inhomogeneous terms at the r.h.s. are suppressed like in (\ref{pha2})
(we denote this approximation by $\cong$),
\be
{\cal H}^{\rm H}_{r,s}  -
\frac{q^{1-2r}}{A}{\cal H}^{\rm H}_{r,s-1}
=  S_r^*\Big(S^*_s - \frac{q}{A} S^*_{s-1}\Big)
%\label{recHu}
\ee
implies
\be
\left(\hat Q_r^2 - \frac{q}{A}\hat P_s\right){\cal H}_{rs} \cong 0
\ee
With operator $\hat P_s$ substituted by its eigenvalue  $z_s$,
we get:
\be
\Sigma^{\rm H}_{(2)}:
\ \ \ \ \ \ \
A\mu_r^2 = qz_s, \ \ \ \ \  A\mu_s^2 = qz_r
\ee
After the change of variables (\ref{chava}) the first equation turns into
\be
\frac{1}{x_r} - \frac{1}{y_s} = 0
\ee
and the variety becomes simply
\be\label{S2}
\Sigma^{\rm H}_{(2)}:
\ \ \ \ \ \ \
x_r-y_s = 0,\ \ \ \ \ \ \
x_s-y_r = 0
\ee
Of course, instead from (\ref{recHu})
one could start from relation
 (\ref{recH}) for the normalized polynomials,
\be
 H^{\rm H}_{r,s}-1 = \frac{q\{q^s\}}{AD_{s-1}}
\Big(q^{-2r}H^{\rm H}_{r,s-1}-1\Big)
\ee

The procedure of getting the varieties can be described by a more formal sequence of steps: take
the difference equation, rewrite them as an operator polynomial of $\hat Q_i$ and $\hat P_i$,
make substitutions
\be
\hat Q^2_i=q^{2i}=\mu_i,
\ \ \ \ \ \ \hat P_i \longrightarrow z_i
\label{limqc1}
\ee
and then put $q=1$. In particular, in the Hopf case one can start with the homogeneous equation of the second order
(\ref{Hh}) and a similar equation for the shift w.r.t. $r$
and immediately obtain the spectral curve as the intersection of products of the unknot curves and (\ref{S2}).

\subsection{Whitehead link}

Again in the limit of small $\hbar$ there is a "trivial" solution
\be
{\rm phase} \ 1\otimes 1 \ {\rm of} \ \cite{AENV}:
\ \ \ \ \ \
\bar{\cal Z}^{\rm W}_{1\otimes 1} = \exp\left(\frac{1}{\hbar}W^{\rm W}_{(1\otimes 1)}(A|x,y)
+ O(\hbar^0)\right), \nn \\
W^{\rm W}_{(1\otimes 1)}(A|x,y) = W^{\rm unknot}(A|x)+ W^{\rm unknot}(A|y)
\ee
it is the same for all links made from the same number of unknots.

What distinguishes different linkings of the two unknots is another $W_{(2)}$ in
another phase, solving the homogeneous equations at the l.h.s. of
(\ref{recuW}) and (\ref{WHcur}).
The corresponding variety $\Sigma^{\rm WH}_{(2)}$ lies in the
intersection of two varieties,  which arise in the double scaling limit
of (\ref{simpleW}) and (\ref{WHcur}) respectively.
The first one is simple:
\be\label{WHsc}
A(\mu_r-\mu_s)(1-z_rz_s)=(A^2\mu_r\mu_s-1)(z_r-z_s)
\ee

This formula can be also obtained (\ref{simpleW}) by the formal procedure described in the previous subsection: make substitutions
\be
r = \frac{\log\mu_r}{\log q^2}\ , \ \ \ \ \ \ s = \frac{\log\mu_s}{\log q^2} \ \ \ \ (\hbox{i.e. }\mu_r=q^{2r},
\ \mu_s=q^{2s}),
\ \ \ \ \ \ \hat P_r \longrightarrow z_r,\ \ \ \ \ \ \hat P_s \longrightarrow z_s,
\label{limqc}
\ee
then put $q=1$ and obtain (\ref{WHsc}). This equation
coincides with the first formula in $V_K(2)$ for the Whitehead link from
\cite[s.7.3]{AENV} after the change of variables
\be\label{cvV}
Q={1\over A^2},\ \ \ \ \ \ \mu_1=\mu_s,\ \ \ \ \ \ \mu_2=\mu_r,\ \ \ \ \ \ \lambda_1={1\over z_sA},
\ \ \ \ \ \ \ \lambda_2={1\over z_rA}
\ee
Similarly, the more complicated second formula (\ref{WHcur})
by the same procedure is converted into a product of
$\frac{A^2\mu_s^2 -1}{A^5\mu_r\mu_s^2}$ and the irreducible equation:
\be
\label{WHcurc}
\mu_s^2\mu_r A^4 +\Big( -\mu_r^2\mu_s^2A^4 + (\mu_r^2-\mu_r\mu_s+2\mu_r-\mu_s)\mu_sA^2 +
(\mu_s-\mu_r)\Big) Az_s+\nn\\
+\Big( (\mu_r\mu_s+\mu_s-2\mu_r)\mu_r\mu_sA^4 +
(\mu_r^2-4\mu_r\mu_s +\mu_s^2 +\mu_r-2\mu_s)A^2 +1 \Big)z_s^2+\nn\\
+\Big( (\mu_r\mu_s-\mu_s^2+2\mu_s-\mu_r)\mu_rA^2 +(\mu_s-\mu_r-1) \Big)Az_s^3+A^2 \mu_r z_s^4=0
\ee
This factorization is immediately seen in (\ref{WHcur}),
because in the limit (\ref{limqc}) all the factors $D_{2s+\ldots}$
become the same, and the only term without such a factor (underlined in (\ref{WHcur}))
is proportional to $\{q^2\}$ and vanishes in the limit.

Eq.(\ref{WHcurc}) coincides
with the second formula in $V_K(2)$ for the Whitehead link from
section 7.3. of \cite{AENV}  after the same change of variables (\ref{cvV}).
Of course, along with (\ref{WHcurc}) there is another equation, obtained by
the substitution $r \leftrightarrow s$.

We discussed here only the l.h.s. of (\ref{WHcur}). As was explained in s.2.4, one can obtain
the fifth order homogeneous equation by acting on (\ref{WHcur}) with the operator $\hat O_r$ canceling the unknot
$S^*_r$. However, in the $q=1$ limit it reduces just to the unknot factor, as in the Hopf case, i.e. comes from
$\Sigma^{\rm WH}_{(1\times 1)}$

\subsection{Borromean rings}

With the same procedure one can immediately generate the spectral variety for the Borromean rings from the recurrent
relations of s.2.5. In particular, the "simple" relations (\ref{simpleB}) lead to
\be
A(\mu_r-\mu_s)(1-z_rz_s)= (A^2\mu_r\mu_s-1)(z_r-z_s)
\label{simpleBV}
\ee
exactly the same, as (\ref{WHsc}).
Two more equations of this type exist for the two other pairs of variables $(r,t)$ and $(s,t)$.

The variety $\Sigma^{\rm B}_{(3)}$
is provided by the intersection of (\ref{simpleBV}) with that
described by the limit (\ref{limqc}) of the "complicated" equation:
\be
\boxed{
\ \
\mu^3\{A\mu\}^3\left(\nu\lambda \{A\mu\}^2(A\mu)^2(1+z^6) +
\sum_{i=1}^5 \sum_{j=-3}^2 \xi^{\rm B}_{ij}\cdot z^iA^{i+2j}\right) = 0
\ \
\label{BorVeq}
}
\ee
Here $(\lambda,\mu,\nu)$ is any permutation of the triple $(\mu_r,\mu_s,\mu_t)$,
and $z = z_\mu$.
The coefficients $\xi_{ij}^{\rm B}$ are
\be
\xi_{1,-1}^{\rm B} =&   -\nu-\lambda+mu-\nu \lambda+\mu \lambda+\nu \mu-\mu^2     \nn \\
\xi_{1,0}^{\rm B} = &  \mu^3 \lambda+\mu^3+\mu^3 \nu-2 \nu \lambda \mu^2-2 \mu^2 \lambda
-2 \nu \mu^2-\mu^2-\mu^2 \lambda^2-\nu^2 \mu^2+\nu \mu
+\nn \\ &+4 \lambda \nu \mu+\nu^2 \lambda \mu
+\lambda^2 \mu+\mu \lambda+\nu \lambda^2 \mu+\nu^2 \mu
-\lambda^2 \nu-\nu \lambda-\nu^2 \lambda     \nn \\
\xi_{1,1}^{\rm B} = & \mu (-\mu^3 \nu \lambda-\mu^3 \nu-\mu^3 \lambda
+\nu \mu^2+4 \nu \lambda \mu^2+\nu^2 \mu^2 \lambda+\mu^2 \lambda^2
+\mu^2 \lambda+\nu \mu^2 \lambda^2
+\nu^2 \mu^2-\lambda^2 \mu  -\nn \\ &-2 \nu \lambda^2 \mu-\nu^2 \mu-2 \lambda \nu \mu
-\nu^2 \lambda^2 \mu-2 \nu^2 \lambda \mu+\nu^2 \lambda^2+\lambda^2 \nu+\nu^2 \lambda)  \nn \\
\xi_{1,2}^{\rm B} =& -\lambda \nu \mu^2 (\mu^2 \lambda+\mu^2+\nu \mu^2-\lambda \nu \mu
-\nu \mu-\mu \lambda+\nu \lambda)
\nn
\ee

\be
\xi_{2,-2}^{\rm B} =& 1+\lambda-2\mu+\nu  \nn \\
\xi_{2,-1}^{\rm B} =&   \nu^2 \lambda+6 \nu \mu^2-2 \mu+\nu^2+\lambda+\lambda^2 \nu
+\nu-2 \lambda^2 \mu-2 \nu^2 \mu+6 \mu^2+3 \nu \lambda-8 \mu \lambda-2 \mu^3
+6 \mu^2 \lambda+\lambda^2-8 \lambda \nu \mu-8 \nu \mu
\nn \\
\xi_{2,0}^{\rm B} =&   -2 \lambda^2 \mu+6 \mu^2 \lambda^2-2 \nu^2 \mu+\lambda^2 \nu
+\nu^2 \lambda+6 \nu^2 \mu^2 \lambda-2 \nu^2 \lambda^2 \mu+6 \nu \mu^2 \lambda^2
 -8 \lambda \nu \mu+26 \nu \lambda \mu^2-8 \nu \lambda^2 \mu -8 \nu^2 \lambda \mu
 - \nn \\ &
-8 \mu^3 \nu \lambda+6 \nu^2 \mu^2+6 \nu \mu^2+6 \mu^2 \lambda+\mu^4 \lambda+\mu^4 \nu
+\nu^2 \lambda^2-8 \mu^3 \nu-2 \mu^3 \lambda^2-8 \mu^3 \lambda-2 \mu^3 \nu^2-2 \mu^3+\mu^4
\nn \\
\xi_{2,1}^{\rm B} =&    \mu (\mu^3 \nu^2 \lambda-2 \nu^2 \mu^2+3 \mu^3 \nu \lambda
-8 \nu \lambda \mu^2-8 \nu \mu^2 \lambda^2+6 \nu^2 \lambda \mu+6 \nu \lambda^2 \mu
-8 \nu^2 \mu^2 \lambda
+ \nn \\ & +6 \nu^2 \lambda^2 \mu-2 \nu^2 \lambda^2+\mu^3 \lambda^2
+\mu^3 \lambda+\mu^3 \nu^2+\mu^3 \lambda^2 \nu-2 \nu^2 \mu^2 \lambda^2
+\mu^3 \nu-2 \mu^2 \lambda^2)
\nn \\
\xi_{2,2}^{\rm B} =& \mu^3\nu\lambda(\mu\lambda-2\nu\lambda+\lambda\nu\mu+\nu\mu)
\nn
\ee

\be
\xi_{3,-3}^{\rm B} =& -1 \nn \\
\xi_{3,-2}^{\rm B} =& 6 \nu \mu+6 \mu-4 \mu^2-1+6 \mu \lambda-2 \lambda-2 \nu \lambda
-2 \nu-\lambda^2-\nu^2 \nn \\
\xi_{3,-1}^{\rm B} =&  -\nu^2 \lambda^2+6 \nu^2 \lambda \mu+6 \mu^3+6 \mu \lambda
-20 \nu \mu^2+6 \lambda^2 \mu-\mu^4-\lambda^2+6 \nu \mu-4 \mu^2 \lambda^2-4 \nu^2 \mu^2
-20 \nu \lambda \mu^2-2 \nu \lambda
-\nn \\ & -2 \nu^2 \lambda+6 \mu^3 \lambda-20 \mu^2 \lambda
-4 \mu^2+6 \mu^3 \nu+6 \nu \lambda^2 \mu+24 \lambda \nu \mu-2 \lambda^2 \nu-\nu^2+6 \nu^2 \mu
\nn \\
\xi_{3,0}^{\rm B} =&   6 \mu^3 \lambda^2-20 \nu \mu^2 \lambda^2+6 \mu^3 \nu^2
-4 \mu^2 \lambda^2+6 \nu^2 \lambda \mu-\mu^4 \nu^2+6 \mu^3 \lambda-4 \nu^2 \mu^2
-20 \nu^2 \mu^2 \lambda-\nu^2 \lambda^2-4 \nu^2 \mu^2 \lambda^2
+ \nn \\ & +6 \nu^2 \lambda^2 \mu-2 \mu^4 \nu
-\mu^4+6 \mu^3 \nu-2 \nu \mu^4 \lambda+6 \mu^3 \lambda^2 \nu
-20 \nu \lambda \mu^2+24 \mu^3 \nu \lambda+6 \mu^3 \nu^2 \lambda-\mu^4 \lambda^2
+6 \nu \lambda^2 \mu-2 \mu^4 \lambda
 \nn \\
\xi_{3,1}^{\rm B} =&   -\mu^2 (-6 \nu^2 \lambda^2 \mu-6 \nu \lambda^2 \mu
+4 \nu^2 \lambda^2+\nu^2 \mu^2+2 \nu \mu^2 \lambda^2+\nu^2 \mu^2 \lambda^2
+2 \nu \lambda \mu^2+2 \nu^2 \mu^2 \lambda+\mu^2 \lambda^2-6 \nu^2 \lambda \mu)     \nn \\
\xi_{3,2}^{\rm B} =&-\mu^4 \nu^2 \lambda^2
\nn
\ee

\be
\xi_{4,-3}^{\rm B} =& 1+\lambda-2 \mu+\nu \nn \\
\xi_{4,-2}^{\rm B} =& \nu^2 \lambda+6 \nu \mu^2-2 \mu+\nu^2+\lambda+\lambda^2 \nu
+\nu-2 \lambda^2 \mu-2 \nu^2 \mu+6 \mu^2+3 \nu \lambda-8 \mu \lambda-2 \mu^3
+6 \mu^2 \lambda+\lambda^2-8 \lambda \nu \mu-8 \nu \mu
\nn \\
\xi_{4,-1}^{\rm B} =&  -2 \lambda^2 \mu+6 \mu^2 \lambda^2-2 \nu^2 \mu
+\lambda^2 \nu+\nu^2 \lambda+6 \nu^2 \mu^2 \lambda-2 \nu^2 \lambda^2 \mu-8 \nu^2 \lambda \mu
+6 \nu \mu^2 \lambda^2-8 \lambda \nu \mu+26 \nu \lambda \mu^2-8 \nu \lambda^2 \mu
- \nn\\ &-8 \mu^3 \nu \lambda+6 \nu^2 \mu^2+6 \nu \mu^2+6 \mu^2 \lambda
+\mu^4 \lambda+\mu^4 \nu+\nu^2 \lambda^2-8 \mu^3 \nu-2 \mu^3 \lambda^2
-8 \mu^3 \lambda-2 \mu^3 \nu^2-2 \mu^3+\mu^4
\nn \\
\xi_{4,0}^{\rm B} =&  \mu (\mu^3 \nu^2 \lambda-2 \nu^2 \mu^2+3 \mu^3 \nu \lambda
-8 \nu \lambda \mu^2-8 \nu \mu^2 \lambda^2+6 \nu^2 \lambda \mu+6 \nu \lambda^2 \mu
-8 \nu^2 \mu^2 \lambda+6 \nu^2 \lambda^2 \mu
-\nn\\ & -2 \nu^2 \lambda^2+\mu^3 \lambda^2
+\mu^3 \lambda+\mu^3 \nu^2+\mu^3 \lambda^2 \nu-2 \nu^2 \mu^2 \lambda^2
+\mu^3 \nu-2 \mu^2 \lambda^2)
\nn\\
\xi_{4,1}^{\rm B} =&  \mu^3 \nu \lambda (\mu \lambda-2 \nu \lambda+\lambda \nu \mu+\nu \mu)
\nn
\ee

\be
\xi_{5,-3}^{\rm B} =& -\nu-\lambda+\mu-\nu \lambda+\mu \lambda+\nu \mu-\mu^2 \nn \\
\xi_{5,-2}^{\rm B} =& \mu^3 \lambda+\mu^3+\mu^3 \nu-2 \nu \lambda \mu^2
-2 \mu^2 \lambda-2 \nu \mu^2 -\mu^2-\mu^2 \lambda^2-\nu^2 \mu^2+\nu \mu
+ \nn \\ & +4 \lambda \nu \mu+\nu^2 \lambda \mu+\lambda^2 \mu+\mu \lambda
+\nu \lambda^2 \mu+\nu^2 \mu-\lambda^2 \nu-\nu \lambda-\nu^2 \lambda \nn \\
\xi_{5,-1}^{\rm B} =&  \mu (-\mu^3 \nu \lambda-\mu^3 \nu-\mu^3 \lambda+\nu \mu^2
+4 \nu \lambda \mu^2+\nu^2 \mu^2 \lambda+\mu^2 \lambda^2+\mu^2 \lambda+\nu \mu^2 \lambda^2
+ \nn\\& +\nu^2 \mu^2-\lambda^2 \mu-2 \nu \lambda^2 \mu-\nu^2 \mu-2 \lambda \nu \mu
-\nu^2 \lambda^2 \mu-2 \nu^2 \lambda \mu+\nu^2 \lambda^2+\lambda^2 \nu+\nu^2 \lambda)   \nn \\
\xi_{5,0}^{\rm B} = &  -\lambda \nu \mu^2 (\mu^2 \lambda+\mu^2+\nu \mu^2-\lambda \nu \mu
-\nu \mu-\mu \lambda+\nu \lambda)
\ee
Making the change of variables (\ref{cvV}):
\be
Q={1\over A^2},\ \ \ \ \ \ \mu_1=\nu,\ \ \ \ \ \ \mu_2=\lambda,\ \ \ \ \ \ \mu_3=\mu,
\ \ \ \ \ \ \lambda_1={1\over z_sA},
\ \ \ \ \ \ \ \lambda_2={1\over z_rA}, \ \ \ \ \ \ l[3]={1\over z_rA},
\ee
 one obtains from (\ref{BorVeq})
and (\ref{simpleBV})
the corresponding equations for $V_K(3)$ for the Borromean rings in s.7.4 of \cite{AENV} and
.

\subsection{Spectral curves}

Equations, derived in the previous subsections describe various hypersurfaces
in the space of $(\mu,z)$-parameters.

In this short subsection we collect all that one can learn about the spectral varieties $V$
of the sequence unknot-Hopf-Whitehead-Borromean rings from the double scaling limit
$\hbar\longrightarrow 0$, $r\longrightarrow \infty$
of the recurrence relations for the HOMFLY polynomials obtained in section 2. These varieties
can be described either as intersections of $\Sigma$'s at different phases, or just as
intersections of all varieties obtained in the limit $q=1$ from all recurrent relations for the given link.

\paragraph{Unknot.} The spectral variety $V_U$ has dimensions 1:
\be
U(x,y) =  xy - x - y + A^2 = 0
\ee

\paragraph{Hopf.} The spectral variety $V_H=U_1\cap U_2\cap H_{12} \cap H_{21}$ has dimension 1:
\be
U(x_1,y_1)=U(x_2,y_2)=0, \nn \\
H(x_1,y_2) = x_1-y_2 = 0, \ \ \ \ \
H(x_2,y_1) = x_2-y_1 = 0
\ee

\paragraph{Whitehead.}   The spectral variety $V_{WH}=U_1\cap U_2 \cap W \cap WH$  has dimension 1:
\be
U(x_1,y_1)=U(x_2,y_2)=0, \nn \\
W = x_1x_2(y_1-y_2) + (x_1-x_2)y_1y_2 - A^2(x_1+y_1-x_2-y_2) = 0\nn \\
WH = (\ref{WHcurc})\ \ \ \ \ \ \ \hbox{with}\ x_{1,2}=\mu_{s,r},\ y_{1,2}={1\over Az_{1,2}}
\ee

\paragraph{Borromean rings.} The spectral variety $V_{B}=U_1\cap U_2 \cap U_3\cap W_{12} \cap W_{13}\cap B$
has dimension 2:
\be
U(x_1,y_1)=U(x_2,y_2)=U(x_3,y_3)=0, \nn \\
W(x_1,x_2,y_1,y_2)= W(x_1,x_3,y_1,y_3)=W(x_2,x_3,y_2,y_3)=0\nn \\
B = (\ref{BorVeq})\ \ \ \ \ \ \ \ \ \hbox{with}\ x_{1,2,3}=\mu_{s,r,t},\ y_{1,2,3}={1\over Az_{1,2,3}}
\ee

\section{Evolution method in application to 2-component links}

Three formulas (\ref{HopfHOMFLY}), (\ref{WhHOMFLY}), (\ref{BoHOMFLY})
clearly form some new interesting sequence,
posing a question, whether an {\it arbitrary} generalized $q$-hypergeometric function
\be
_uF_v\left( {{a_1,\ldots,a_u}\atop {b_1,\ldots,b_v}} \Big|\ z\right) =
\sum_k q^{Q(k)}\, \frac{(a_1)_k\ldots (a_u)_k}{(b_1)_k\ldots(b_v)_k}\frac{z^k}{[k]!}
\ee
with $(a)_k = \frac{(a+k)!}{(a)!}$ and certain quadratic form $Q(k)$
{\it can be} a HOMFLY polynomial for {\it some} link.

\bigskip

An alternative interesting question is how much the knot polynomials
for other simple links {\it deviate} from this simple structure.
In this section we provide results about the link family,
to which the Hopf and Whitehead links naturally belong
and study the evolution {\it a la} \cite{DMMSS,evo}
along this family (see Figure 2).

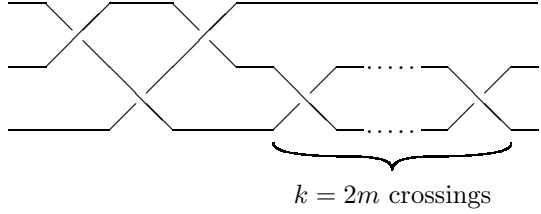
\begin{figure}
\begin{picture}(360,70)(-100,-37)
\put(0,0){\line(1,0){38}}
\put(0,24){\line(1,0){14}}
\put(0,48){\line(1,0){14}}
\put(14,24){\line(1,1){24}}
\put(14,48){\line(1,-1){10}}
\put(28,34){\line(1,-1){34}}
\put(38,0){\line(1,1){10}}
\put(52,14){\line(1,1){34}}
\put(38,48){\line(1,0){24}}
\put(62,48){\line(1,-1){10}}
\put(62,0){\line(1,0){38}}
\put(76,34){\line(1,-1){10}}
\put(86,24){\line(1,0){14}}
\put(100,0){\line(1,1){10}}
\put(100,24){\line(1,-1){24}}
\put(114,14){\line(1,1){10}}
\put(124,0){\line(1,0){10}}
\put(124,24){\line(1,0){10}}
\multiput(135,-1)(4,0){5}{.}
\multiput(135,23)(4,0){5}{.}
\put(156,0){\line(1,0){10}}
\put(156,24){\line(1,0){10}}
\put(166,0){\line(1,1){10}}
\put(166,24){\line(1,-1){24}}
\put(180,14){\line(1,1){10}}
\put(190,0){\line(1,0){10}}
\put(190,24){\line(1,0){10}}
\put(86,48){\line(1,0){114}}
\qbezier(100,-5)(100,-10)(130,-10)
\qbezier(190,-5)(190,-10)(160,-10)
\qbezier(130,-10)(145,-10)(145,-15)
\qbezier(160,-10)(145,-10)(145,-15)
\put(145,-25){\makebox(0,0)[cc]{$k=2m$ crossings}}
\end{picture}
\caption{{\footnotesize
This figure describes the braid for the evolution within a
family of $2$-component $3$-strand links, which includes the Hopf link
($k=0$) and the Whitehead link ($k=2$).
When $k=0$ the link described by this braid has three crossings instead
of two for the Hopf link. To get the latter one the first Reidemeister
move should be applied to the lower crossing.}}
\end{figure}

Thus, we consider the
family of the $L_{2k+1}=[k,1;2]$ links (it is also easy to extend to $[k,l;2]$,
however increasing the length of braid with counter-oriented strands from
$2$ to $m$ is more problematic).
We present the results in the form of the differential expansion
of \cite{Art}, generalizing the dream-like formulas of \cite{IMMMfe}
for the figure eight knot.

\paragraph{The series $L_{2k+1}$ in representation $[r]\otimes [1]$}.
These are all links of the two unknots, thus the answer is symmetric
under the permutation of $r$ and $s$.
In the following list we refer also to notation from the link classification
in \cite{katlas}.
Note that $k$ in this family is always even, $k=2m$.
\be
{\rm Hopf:} & H_{[r],[1]}^{k=0} = 1 - { \frac{1}{Aq^{r-1}}\frac{\{q^r\}\{q\}}{D_0}  }, \nn \\
{\rm Whitehead:} & H_{[r],[1]}^{k=2} = 1 + \frac{\{q^r\}\{q\}D_rD_{-1}}{AD_0}, \nn \\
{\rm L7a6:} & H_{[r],[1]}^{k=4} = 1 + {  Aq^{r-1}\frac{\{q^r\}\{q\}}{D_0}  }
+ \frac{1+q^{2r+2}}{q^2}\frac{\{q^r\}\{q\}D_rD_{-1}}{AD_0}, \nn \\
{\rm L9a36:} & H_{[r],[1]}^{k=6} = 1 + Aq^{r-1}(1+A^2q^{2r-2})\frac{\{q^r\}\{q\}}{D_0}
+ \left(\frac{1+q^{2r+2}+q^{4r+4}}{q^4}- A^2q^{2r-2}\right)\frac{\{q^r\}\{q\}D_rD_{-1}}{AD_0}, \nn \\
{\rm L11a360:} & H_{[r],[1]}^{k=8} = 1 + Aq^{r-1}(1+A^2q^{2r-2}+A^4q^{4r-4})\frac{\{q^r\}\{q\}}{D_0}
+ \nn \\  &
+ \left(\frac{1+q^{2r+2}+q^{4r+4}+q^{6r+6}}{q^6}
-A^2q^{2r-2}(q^{-2}+1+q^{2r})-A^4q^{4r-4}\right)\frac{\{q^r\}\{q\}D_rD_{-1}}{AD_0}, \nn \\
\ldots
\ee
Clearly,
\be
H_{[r],[1]}^{k} = 1 + Aq^{r-1}\ \frac{1-(Aq^{r-1})^{k-2}}{1-(Aq^{r-1})^2}\frac{\{q^r\}\{q\}}{D_0}
+ \frac{1-q^{k(r+1)}}{q^{k-2}(1-q^{2(r+1)})}\frac{\{q^r\}\{q\}D_rD_{-1}}{AD_0} - \nn \\
-\left( \sum_{j=0}^{k/2-3}A^{2j+2}\sum_{i=0}^{k/2-3-j} q^{(k-4)(r-1)+2(k/2-3-j)-(2r+1)i}[i+1]_q\right)
\frac{\{q^r\}\{q\}D_rD_{-1}}{AD_0}
\label{Hr1naive}
\ee
Looking at these formulas one can get an impression that the powers
of $A$ increase with $r$, however this is not true: they cancel between
the two structures in this formula.
This is getting clear from an alternative expression (\ref{Hr1fromevo}) below.

In fact,
one can consider these formulas from the point of view of
the evolution method of \cite{DMMSS} and \cite{evo}.
The product of representations $[r]\otimes [1] = [r+1] + [r,1]$,
and
\be
\varkappa_{[r+1]} - \varkappa_{[r]}-\varkappa_{[1]} =
\frac{r(r+1)}{2} - \frac{r(r-1)}{2} - 0 = r, \nn \\
\varkappa_{[r,1]} - \varkappa_{[r]}-\varkappa_{[1]} =
\frac{(r-2)(r+1)}{2} - \frac{r(r-1)}{2} - 0 = -1,
\ee
thus one expects that
\be
H_{[r],[1]}^{k} =  aq^{kr} + bq^{-k}
\ee
with some parameters $a$ and $b$, which can depend on $A$ and $q$,
but not on $k$.
This is indeed true, with
\be
a= \frac{\{Aq^r\}}{A^2q^{2r-2}[r+1]_q\{A\}}
(q^{2r-2}A^2-q^{2r}+1), \nn \\
b= \frac{[r]_q\{A/q\}}{A^2q^{2r}[r+1]_q\{A\}}
(q^{2r}A^2+q^2-1),
\ee
i.e.
\be
\boxed{\
H_{[r],[1]}^{k} =
\frac{(q^{2r-2}A^2-q^{2r}+1)\,\{q\}D_r\, q^{2+kr}\ +\
(q^{2r}A^2+q^2-1)\,\{q^r\}D_{-1}\,q^{-k}}{A^2q^{2r}\{q^{r+1}\}D_0}
\ }
\label{Hr1fromevo}
\ee
This time it is clear that the powers of $A$ are limited,
however, instead the differential structure of (\ref{Hr1naive})
is obscure.
An alternative formulation, where the both properties are
transparent is as follows:
\be
H_{[r],[1]}^{k} = 1 + \frac{\Big(f_+A^2 + f_0 + f_-A^{-2}\Big)\{q^r\}\{q\}}{AD_0}
\label{Hr1f}
\ee
with
\be
f_+=q^{r-1}\left(\frac{q^{2mr}-1}{q^{2r}-1} + q^{-2}\frac{q^{2(m-1)r}-1}{q^{2r}-1} +
\ldots + q^{-2(m-1)}\frac{q^{2r}-1}{q^{2r}-1}\right), \nn \\
f_0 = - q^{(m-1)(r-1)}\left(q^{m(r+1)}+q^{-m(r+1)} + 2\,\frac{\{q^{(m-1)(r+1)}\}}{\{q^{r+1}\}}
+q^{(m-1)+(m-2)r}\cdot\sum_{j=1}^{m-2}\ [j]\cdot q^{-(2r+1)j}\right), \nn \\
f_- = q^{(m-2)(r-1)}\frac{\{q^{m(r+1)}\}}{\{q^{r+1}\}}
\ \ \ \ \ \ \ \ \ \ \ \ \ \ \ \ \ \ \ \ \ \ \ \ \ \ \ \ \ \ \ \ \ \ \ \ \ \ \ \ \ \ \ \ \
\ee
For $m\leq 3$ the standard summation rule is implied in the expression for $f_0$: e.g.
$\sum_{j=1}^{-2} g(j) = -g(-1) - g(0)$.
A generalization of (\ref{Hr1f}) to arbitrary $s$  is
\be
H_{rs}^{k} =1+ \sum_{k=1}^{{\rm min}(r,s)}
\prod_{j=0}^{k-1} \left(\frac{\{q^{r-j}\}\{q^{s-j}\}}{AD_j}\right)
\left(\sum_{i=-k}^{k} f_{-k+2j}^{(rs)}A^{2i}\right)
\ee
and the coefficients $f$ remain to be determined.

\bigskip

One way to do this is to return to the evolution method.
In generic symmetric  representations $[r]$ and $[s]$
\be
H^k_{r,s} = \sum_{p=0}^{{\rm min}(r,s)}  c^{(p)}_{r,s} q^{k\big((r-p)(s-p) - p\big)}
\label{evoHopf}
\ee
and for $s=1$
\be
c^{(0)}_{r,1}=
\frac{\{q\}D_r}{A^2q^{2(r-1)}\{q^{r+1}\}D_0}\Big(q^{2r-2}A^2-q^{2r}+1\Big) \nn \\
c^{(1)}_{r,1}= \frac{\{q^r\}D_{-1}}{A^2q^{2r}\{q^{r+1}\}D_0}\Big(q^{2r}A^2+q^2-1\Big)
\ee
while for $s=2$
\be
c^{(0)}_{r,2}=
\frac{\{q\}\{q^2\}D_rD_{r+1}}{A^4q^{4r-4}\{q^{r+1}\}\{q^{r+2}\}D_0D_1}
\Big(q^{4(r-1)}A^4-[2]q^{3(r-1)}\{q^r\}A^2 + q^{2r-1}\{q^r\}\{q^{r-1}\}\Big) \nn \\
c^{(1)}_{r,2}=
\frac{\{q^2\}D_rD_{-1}}{A^4q^{4r}\{q^{r+2}\}D_0D_1}
\Big(q^{4r}A^4- q^{2r}A^2(q^r\{q^r\}-\{q^2\})-q^{r+1}\{q^2\}\{q^{r-1}\}
\Big)    \nn\\
c^{(2)}_{r,2}=
\frac{\{q^{r-1}\}D_{-1}}{A^4q^{4r+2}\{q^{r+1}\}D_1}
\Big(q^{4r+2}A^4 + q^{2r+2}\{q^2\}A^2 + q^{3}\{q^2\}\{q\}\Big)
\ee
Clearly, with the obvious notation $D_r!\equiv  \prod_{j=0}^r D_j$
(note that the product starts from $j=0$ and includes $r+1$ factors,
also note that according to this definition $D_{-1}!=1$ and $D_{-2}!=1/D_{-1}$),
\be
c^{(0)}_{r,s}=
\frac{[r]!\,[s]!}{[r+s]!}\frac{D_{r+s-1}!}{D_{r-1}!\,D_{s-1}!}
\sum_{j=0}^{{\rm min}(r,s)} \frac{(-)^jq^{\frac{j(j+5)}{2}-j(r+s)}\,\{q\}^j }{A^{2j}}
\frac{[r]!\,[s]!}{[r-j]!\,[s-j]!\,[j]!}
\ee
and in general
\be
c^{(p)}_{r,s} =
\frac{[r]![s]![r+s+1-2p]}{[p]![r+s+1-p]!}\cdot
\frac{D_{r+s-p-1}!  }{D_{r-1}!D_{s-1}!}\prod_{i=1}^{p} D_{i-2}\cdot
\left( 1+  \sum_{j=1}^{{\rm min}(r,s)} \frac{\sigma_{r,s}^{(p|j)}}{A^{2j}} \right)
\ee
Note that the factor $[r+s+1-2p]$ in the numerator
is just a single quantum number, not a factorial.

The matrices $\sigma$ are symmetric under the permutation of $r$ and $s$,
the first two of them are:

\be
\sigma^{(0|j)}_{rs} = (-)^jq^{ \frac{1}{2}{j(j+5)}-j(r+s)}\,
\prod_{i=0}^{j-1}\frac{\{q^{r-i}\}\{q^{s-i}\}}{\{q^{i+1}\}}
\nn \\
\sigma^{(1|j)}_{rs} = (-)^{j} q^{\frac{1}{2}{j(j+3)} -j(r+s)}
\,\underbrace{\frac{\{q^{s-j}\}\{q^r\}-q^{-r}\{q^j\}\{q^s\}}{\{q\}}}\,
\prod_{i=1}^{j-1}\frac{\{q^{r-i}\}\{q^{s-i}\}}{\{q^{i+1}\}}
\ee
The underbraced ratio respects $r\leftrightarrow s$ symmetry,
because it is equal to
\be
{\frac{q^{r+s-j} -q^j(q^{r-s}+q^{s-r})+(2q^{j}-q^{-j})q^{-r-s}}{\{q\}}}
\ee

In general
\be
\sigma^{(p,j)}_{r,s}=
\displaystyle{{(-)^{p+j}\cdot q^{(p-1)(p-2j)}\cdot q^{\frac{j(j+1)}{2}-j(r+s)}\over [j]!}}\cdot
\{q\}^j\cdot \nn \\
\cdot
\sum_{a,b=0}^p\left(\frac{[p]!}{[a]![b]![p-a-b]!}\, (-)^{a+b}\cdot q^{(a+b)(j+1-p)}\cdot q^{ab}
\cdot\left( \prod_{i=0}^{j-1}\, [r-a-i][s-b-i]\right)\right)
\ee

To apply Zeilberger's programs and obtain recursion relations
one should begin with the differential-hierarchy
analysis {\it a la} \cite{Art} to convert the answers to
the hypergeometric form, a generalization of (\ref{HopfHOMFLY})-(\ref{BoHOMFLY}).
This will be done elsewhere.

\section{Evolution of three-component Borromean rings}

\begin{figure}
\begin{picture}(360,90)(-130,-37)
\put(-48,0){\line(1,0){38}}
\put(-48,24){\line(1,0){14}}
\put(-48,48){\line(1,0){14}}
\put(-34,24){\line(1,1){24}}
\put(-34,48){\line(1,-1){10}}
\put(-20,34){\line(1,-1){34}}
\put(-10,48){\line(1,0){24}}
\put(14,0){\line(1,0){24}}
\put(-10,0){\line(1,1){10}}
\put(4,14){\line(1,1){34}}
\put(14,48){\line(1,-1){10}}
\put(28,34){\line(1,-1){34}}
\put(38,0){\line(1,1){10}}
\put(52,14){\line(1,1){34}}
\put(38,48){\line(1,0){24}}
\put(62,48){\line(1,-1){10}}
\put(62,0){\line(1,0){24}}
\put(76,34){\line(1,-1){34}}
\put(86,0){\line(1,1){10}}
\put(100,14){\line(1,1){10}}
\put(110,0){\line(1,0){14}}
\put(110,24){\line(1,0){14}}
\put(124,0){\line(1,1){10}}
\put(124,24){\line(1,-1){24}}
\put(138,14){\line(1,1){10}}
\put(148,0){\line(1,0){10}}
\put(148,24){\line(1,0){10}}
\multiput(159,-1)(4,0){5}{.}
\multiput(159,23)(4,0){5}{.}
\put(180,0){\line(1,0){10}}
\put(180,24){\line(1,0){10}}
\put(190,0){\line(1,1){10}}
\put(190,24){\line(1,-1){24}}
\put(204,14){\line(1,1){10}}
\put(214,0){\line(1,0){10}}
\put(214,24){\line(1,0){10}}
\put(86,48){\line(1,0){138}}
\qbezier(124,-5)(124,-10)(154,-10)
\qbezier(214,-5)(214,-10)(184,-10)
\qbezier(154,-10)(169,-10)(169,-15)
\qbezier(184,-10)(169,-10)(169,-15)
\put(169,-25){\makebox(0,0)[cc]{$k=2m$ crossings}}
\put(234,-5){\makebox{$r$}}
\put(234,22){\makebox{$s$}}
\put(234,45){\makebox{$t$}}\label{f3}
\end{picture}
\caption{{\footnotesize This figure describes the braid for
the evolution within the family of $3$-component $3$-strand links,
which includes the Borromean rings ($k=0$)
and link $L8n5$ (for $k=-4$).}}
\end{figure}
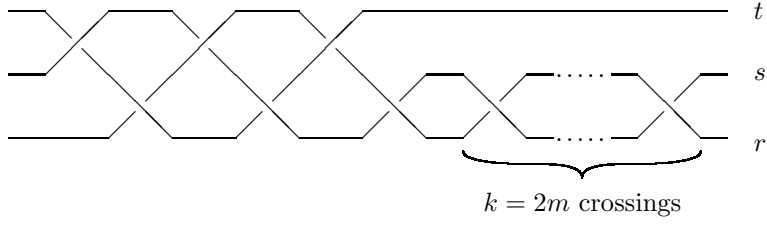

This time we consider the evolution in parameter $k=2m$
defined in picture \ref{f3}.
The $k=0$ member of the family is the Borromean link,
while $L8n5$ also considered in \cite{AENV}, arises at $k=-4$.

The evolution formula
this time is
\be
H^k_{r,s,t} = \sum_{p=0}^{{\rm min}(r,s,t)}
c^{(p)}_{r,s,t} q^{{k\big((r-p)(s-p) - p\big)}}
\label{evoBor}
\ee
and it should be analyzed just in the same way as (\ref{evoHopf})
was in the previous section.

Here we just mention the first non-trivial result:
\be
c^{(0)}_{1,s,t} =\frac{1}{[s+1]}\frac{D_s}{D_0^2D_1}\Big(D_0D_1-
q^{s+1}D_{-1}D_t[s][t]\{q\}^3\Big),
\nn \\
c^{(1)}_{1,s,t} = \frac{[s]}{[s+1]}\frac{D_{-1}}{D_0^2D_1}
\Big(D_0D_1+q^{-s-1}D_sD_t[t]\{q\}^3\Big)
\label{cBor01}
\ee
sufficient to define $H_{1,s,t}^k$ for the entire
family,when one   of the three representations is
fundamental, while the other two are arbitrary.
Note that for $k\neq 0$ there is a symmetry only between two
of the three representations (between $r$ and $s$),
thus there is no symmetry between $s$ and $t$ in (\ref{cBor01}).

\section*{Acknowledgements}

A.Morozov and And.Morozov are indebted for hospitality to the GGI,
where part of this work was done. Our work is partly supported by
Ministry of Education and Science of the Russian Federation under
contract 8606, by the Brazil National Counsel of Scientific and
Technological Development (A.Mor.), by the Laboratory of Quantum Topology of
Chelyabinsk State University (Russian Federation government grant 14.Z50.31.0020) (And.Mor.).
by NSh-3349.2012.2 (A.Mir.,
A.Mor.,And.Mor.), by RFBR grants 12-01-00482 (S.A.), 13-02-00457 (A.Mir.), 13-02-00478
(A.Mor.), 11-01-00962 (And.Mor.), by joint grants 12-02-92108-Yaf (A.Mir., A.Mor.),
13-02-91371-ST (A.Mir., A.Mor., And.Mor.), 14-01-93004-Viet (A.Mir.,
A.Mor.), 14-02-92009-NNS (A.Mir., A.Mor.), by leading young scientific groups RFBR 12-01-31385
mol$\_$a$\_$ved (S.A.), 12-02-31078
mol$\_$a$\_$ved  and by D.~Zimin's ``Dynasty''
foundation (And.Mor.). The author is partially supported by

\end{document}